\documentclass[a4paper,fleqn]{cas-sc}
\usepackage{diagbox}
\usepackage[authoryear,sort&compress]{natbib}
\usepackage{wrapfig}
\usepackage[ruled,vlined]{algorithm2e}
\usepackage{algpseudocode} 
\usepackage{graphicx}
\graphicspath{ {Figures/} }
\usepackage{caption}

\usepackage{algpseudocode}
\usepackage{subcaption}
\usepackage{etoolbox} 

\makeatletter
\patchcmd{\@algocf@start}{\vskip\ALG@thistlm}{\vskip 0.5ex}{}{}
\patchcmd{\@algocf@start}{\vskip\ALG@thistlm}{\vskip 0.5ex}{}{}
\makeatother

\def\tsc#1{\csdef{#1}{\textsc{\lowercase{#1}}\xspace}}
\tsc{WGM}
\tsc{QE}
\tsc{EP}
\tsc{PMS}
\tsc{BEC}
\tsc{DE}

\begin{document}
\let\WriteBookmarks\relax
\def\floatpagepagefraction{1}
\def\textpagefraction{.001}
\shorttitle{}

\title [mode = title]{Hybrid Deep Learning Framework for Classification of Kidney CT Images: Diagnosis of Stones, Cysts, and Tumors}                      
\author[1,2]{Kiran Sharma}\corref{cor2}
\ead{kiran.sharma@bmu.edu.in}

%
%
\author[1,2]{Ziya Uddin}\corref{cor1}
\ead{ziya.uddin@bmu.edu.in}
%
\author[1]{Adarsh Wadal}
\author[1]{Dhruv Gupta}

\address{School of Engineering \& Technology, BML Munjal University, Gurugram, Haryana-122413, India }
\address{Center for Advanced Data and Computational Science, BML Munjal University, Gurugram, Haryana-122413, India }
\cortext[cor1]{Corresponding author1}
\cortext[cor2]{Corresponding author2}
\begin{abstract}
Medical image classification is a vital research area that utilizes advanced computational techniques to improve disease diagnosis and treatment planning. Deep learning models, especially Convolutional Neural Networks (CNNs), have transformed this field by providing automated and precise analysis of complex medical images. This study introduces a hybrid deep learning model that integrates a pre-trained ResNet101 with a custom CNN to classify kidney CT images into four categories: normal, stone, cyst, and tumor. The proposed model leverages feature fusion to enhance classification accuracy, achieving 99.73\% training accuracy and 100\% testing accuracy. Using a dataset of 12,446 CT images and advanced feature mapping techniques, the hybrid CNN model outperforms standalone ResNet101. This architecture delivers a robust and efficient solution for automated kidney disease diagnosis, providing improved precision, recall, and reduced testing time, making it highly suitable for clinical applications.
\end{abstract}

\maketitle
\section{Introduction}

CT (computed tomography) imaging is an integral part of the diagnostic workup for renal diseases. The necessity of classifying kidney conditions as normal, stone, cyst, or tumor from CT images is to enable relevant medical intervention. This process is called multi-modal classification, where using CT imaging techniques allows for the effective differentiation of these conditions. Multi-modal classification is of primary importance due to its potential to enhance diagnostic accuracy and improve patient outcomes.

Correct distinction between normal kidney, stone, cyst, and tumor is crucial for treatment planning, as each disorder requires a tailored procedure. For instance, a simple cyst may need little or no treatment, while a malignant tumor often requires aggressive measures like surgery or chemotherapy. Early and accurate classification can have a profound effect on the treatment regimen and prognosis of renal diseases. Medical imaging and analysis is a highly progressive field, particularly because of advancements in machine learning.

 \cite{patro2023kronecker} proposed a novel variant of convolutional techniques, known as the Kronecker convolution method, for better feature extraction from CT images using their models for classification through improving classifiers. This decreases duplication of feature maps, thus bettering CNN performance for detecting kidney stones by capturing more salient features. Additionally, \cite{wu2020automated} have exhibited the ability to combine CNNs through feature fusion of more than one neural network, markedly enhancing the identification and categorization of renal pathologies. The pioneering study developed a multi-features CNN fusion model for detection of multiple kidney anomalies from ultrasound images.
 
In addition, deep learning algorithms have been developed to detect subtle differences in kidney tissues, allowing the identification of small kidney stones or early-stage tumors \citep{yildirim2021deep}. For example, the ExDark19 model proposed by \cite{baygin2022exemplar} applies transfer learning and feature selection techniques, achieving the model accuracy of 99\%  in kidney stone detection. Similarly, the model proposed by \cite{caglayan2022Deep} achieved an accuracy rate of 99. 1\% to detect kidney stones. \cite{Hossain2023Kidney} evaluated EANet with accuracy 83.65\%, ResNet50 with accuracy 87.92\%, and a CNN model with accuracy 98.66\%. Furthermore, \cite{gulhane2024integrative} introduced an integrative approach that utilizes an improved deep neural network (DNN) architecture for the efficient detection of kidney stones. By combining traditional machine learning techniques with DNNs, the approach improves pattern recognition in datasets. \cite{elhoseny2019intelligent} proposed a diagnostic prediction and classification system for chronic kidney disease (CKD) that integrates density-based feature selection (DFS) with ant colony optimization (ACO). By eliminating irrelevant data before classification, the system improves the accuracy and efficiency of CKD detection and management. 

Despite significant advances, challenges remain in the multi-modal classification of kidney conditions. Variability in image quality and the complexity of kidney pathologies often result in inconsistent classification outcomes. Furthermore, there is a pressing need for larger and more diverse datasets to train models, ensuring robustness and generalizability across populations and clinical settings.
\section{Related work}

Recent advances in deep learning methods have greatly improved image classification in numerous domains. These advanced methods have also improved the classification of medical images in disease diagnosis such as kidney, retinal, cancer, etc.  CNN models demonstrated high accuracy, precision, and recall rates, making them promising tools for the early detection and classification of such diseases. There is sufficient literature on multi-modal analysis of kidney abnormalities.

\cite{rajora2024advancing} explored CNNs for multi-modal classification of kidney diseases. They proposed a combined ensemble model, VGG16 and InceptionV3, achieving a model accuracy of 97.8\%. Similarly, \cite{ali2023automated} developed a custom CNN model that demonstrated high validation accuracy and precision in kidney disease classification, highlighting its robustness and efficiency in managing complex classifications. They introduced a hybrid CNN architecture with feature fusion and ensemble classification, achieving an accuracy of 94.5\%.

The attention-based CNN model with dual attention mechanisms proposed by \cite{chauhan2024attention} significantly enhanced feature extraction from CT images, thereby improving diagnostic performance. Along similar lines, \cite{sri2023deep} proposed a deep ensemble model for kidney disease detection, achieving high precision and low loss, showcasing the effectiveness of combining multiple deep learning models for improved diagnostic results. Furthermore, \cite{bindumadavi2023deepkidney} introduced the ``DeepKidney'' model, which utilizes CNNs for classifying kidney diseases, including stones, cysts, and tumors, demonstrating notable advancements in diagnostic accuracy and efficiency.

\cite{sakib2023kidney} introduced a CNN model which was a combination of customized CNN and ResNet50 and achieved an accuracy of 98.66\%. This study underscores the effectiveness of CNN-specific methodologies combined with transfer learning to enhance classification performance. Similarly, \cite{singh2023precision} used the EfficientNet-B3 model, fine-tuned on a large CT kidney image dataset, achieving a classification accuracy of 95.77\%. Their approach highlights the efficiency of the model in processing medical images and delivering reliable diagnostic results. \cite{sharma2024revolutionizing} designed a robust CNN-based framework that achieved an impressive precision of 99. 88\% in the classification of multiclass kidney disease. This framework established a new standard for precision and reliability in detecting various kidney conditions.

Chowdhury et al. \cite{chowdhury2023mutual} utilized a mutual learning algorithm comprising multiple CNN and SVM models to enhance diagnostic accuracy while addressing data limitations and privacy concerns. This method improved learning efficiency through dynamic teacher-student interactions. Similarly, the YOLOv8 model was applied by \cite{pande2024multi} for the detection of kidney cysts, stones, and tumors with high accuracy, highlighting the potential of real-time detection systems in supporting rapid diagnosis and treatment planning in clinical settings.

Two studies emphasize the effectiveness of VGG16 and InceptionV3 models in detecting kidney diseases. \cite{prasher2024vgg16} achieved a training and validation accuracy of 99.7\% using the VGG16 model, demonstrating its strong ability to accurately classify kidney disorders. Meanwhile, \cite{anand2024transfer} modified the InceptionV3 model, achieving 96.52\% accuracy in identifying conditions such as cysts, stones, and tumors. The ability of the VGG16 model to generalize from training to unseen data underscores its real-world applicability \citep{prasher2024vgg16}. Both studies highlight that transfer learning models outperform traditional diagnostic methods, with their high precision showcasing their potential to aid in the early and accurate diagnosis of kidney diseases.

\begin{table}[!ht]
\centering
\caption{Comparative analysis of related literature.}
\resizebox{\textwidth}{!}{  
\begin{tabular}{lllcl}
\hline
\textbf{\begin{tabular}[c]{@{}l@{}}Paper\\ Refernce\end{tabular}}& \multicolumn{1}{l}{\textbf{Data Distribution}}            & \multicolumn{1}{l}{\textbf{Methodology}}        & \multicolumn{1}{c}{\textbf{\begin{tabular}[c]{@{}c@{}}Accuracy \\ (\%)\end{tabular}}} & \multicolumn{1}{l}{\textbf{Limitations}}     \\ \hline

\cite{sharma2024revolutionizing}  & \cellcolor[HTML]{FFFFFF}{\color[HTML]{1F1F1F} \begin{tabular}[c]{@{}l@{}}Normal: 5,077, Stone: 1,377, \\ tumor: 2283, and Cyst: 3709\end{tabular}} & Fine-tuned InceptionV3    & 96.52            & \begin{tabular}[c]{@{}l@{}}Lower performance in tumor detection \\ as compared to other classes.\end{tabular}    \\  \hline

\cite{prasher2024vgg16}  & \begin{tabular}[c]{@{}l@{}}Normal: 4500, Stone: 1000, \\ tumor: 3000, and Cyst: 2000\end{tabular}  & \begin{tabular}[c]{@{}l@{}}Transfer learning with \\ pre-trained VGG16 .\end{tabular}    & 98.8     & \begin{tabular}[c]{@{}l@{}}Over-reliance on pre-trained weights; \\ lower performance on imbalanced datasets.\end{tabular}   \\ \hline

\cite{Hossain2023Kidney}  & \begin{tabular}[c]{@{}l@{}}Normal: 5,077, Stone: 1,377,  \\ tumor: 2283, and Cyst: 3709\end{tabular}  & \cellcolor[HTML]{FFFFFF}custom CNN   & \cellcolor[HTML]{FFFFFF}99.92       & \begin{tabular}[c]{@{}l@{}}Watershed method struggle with over-segmentation, \\ noise sensitivity, and detecting thin or low-contrast areas.\end{tabular} \\ \hline

\cite{chowdhury2023mutual} & \cellcolor[HTML]{FFFFFF}{\color[HTML]{1F1F1F} \begin{tabular}[c]{@{}l@{}}Normal: 5,077, Stone: 1,377, \\ tumor: 2283, and Cyst: 3709\end{tabular}} & \begin{tabular}[c]{@{}l@{}}CNN and SVM through \\ teacher-student interaction\end{tabular} & \begin{tabular}[c]{@{}c@{}}85.76;\\ 74.04\end{tabular}   & \begin{tabular}[c]{@{}l@{}}The model's generalization to new \\ datasets remains a challenge.\end{tabular}    \\  \hline

\cite{bindumadavi2023deepkidney}  & Specific counts not mentioned  & CNN     & 98.91        & \begin{tabular}[c]{@{}l@{}}Lower performance in tumor detection \\ as compared to other classes.\end{tabular}   \\ \hline

\cite{chauhan2024attention}      & \begin{tabular}[c]{@{}l@{}}Includes pre-processed \\ kidney ailment images\end{tabular}   & Attention-based DNN    & 95.5     & \begin{tabular}[c]{@{}l@{}}Scaling to larger datasets poses \\ computational challenges.\end{tabular}    \\ \hline

\cite{singh2023precision}    & \begin{tabular}[c]{@{}l@{}}1,500 images: 500 each \\ for cyst, tumor, and stone\end{tabular} & Fine-tuned EfficientNet-B3    & 95.77  & \begin{tabular}[c]{@{}l@{}}Challenges in distinguishing overlapping \\ features between tumor and stone cases.\end{tabular}   \\ \hline

\cite{rajora2024advancing}   & Dataset of CT images    & \begin{tabular}[c]{@{}l@{}}Ensemble model combining \\ VGG16 and InceptionV3\end{tabular}     & 97.8 & \begin{tabular}[c]{@{}l@{}}High training time due to complex architecture; \\ requires significant computational resources.\end{tabular} \\ \hline

\cite{sri2023deep}   & \begin{tabular}[c]{@{}l@{}}Normal: 4000, Stone: 2500, \\ tumor: 1500, and Cyst: 3000\end{tabular}   & \begin{tabular}[c]{@{}l@{}}Pre-trained VGG16 and fine \\ tunned InceptionV3\end{tabular}   & 95.7    & \begin{tabular}[c]{@{}l@{}}Normalization dependencies limit adaptability \\ to real-world noisy datasets \end{tabular}  \\ \hline

\cite{ali2023automated}    & \begin{tabular}[c]{@{}l@{}}Balanced dataset of 2,000 for \\ each  class\end{tabular}  & \begin{tabular}[c]{@{}l@{}}Hybrid CNN model \\ integrating feature fusion \\ and ensemble classification\end{tabular} & 94.5       & \begin{tabular}[c]{@{}l@{}}Requires additional optimization for handling \\ cross-modality image datasets.\end{tabular}       \\ \hline                           
\end{tabular}
}
\label{tab:literature}
\end{table}
\subsection{Motivation of study}
Medical image classification is a key area of research, with numerous automated systems developed for accurate disease detection. Deep learning models have significantly contributed to the achievement of high performance in this field. Convolutional Neural Networks (CNNs) are widely used in biomedical computer vision tasks \citep{krizhevsky2012imagenet}. Most existing studies focus on customizing single pre-trained models, such as custom CNNs, attention-based DNN, fine-tuned EfficientNet-B3, SVM models,pre-trained VGG16, fine-tunned InceptionV3, etc. as summarized in Table~\ref{tab:literature}. These proposed models suffers from certain limitations. To overcome some of the limitations, the proposed hybrid CNN model offers an effective feature-engineering approach, combining pre-trained ResNet101 and a custom CNN through feature fusion. This model achieves high accuracy in classifying kidney CT images. The classification process involves three key steps: first, feature extraction and classification using the pre-trained ResNet101; second, feature extraction with the custom CNN; and finally, feature fusion to classify images into categories such as normal, stone, cyst, and tumor.
\subsection{Research objectives}
The objectives of the study are two-folds:

\begin{itemize}
    \item To develop a hybrid CNN model to classify the kidney as normal, stone, cyst, and tumor with improved accuracy.
    \item To compare the classification performance of the proposed hybrid CNN with base models: CNN and ResNet101. 
\end{itemize}

\section{Methodology}
\subsection{Data description and pre-processing}

The ultrasound or CT scan data set of kidneys with kidney stones, tumor, or cyst was sourced from Kaggle (\url{https://www.kaggle.com/datasets/srinivasbece/kindey-stone-dataset-splitted}). The data set was intentionally designed to encompass various types of kidney disease, taking into account the variability in size, location, and composition. This extensive data set was designed to enhance the accuracy and robustness of the system, allowing effective identification and classification of kidney diseases with a comprehensive understanding of clinical manifestations. The data of the images of kidney diseases are classified into four categories, namely: Normal, Stone, Cyst, and Tumor. Table~\ref{tab:DataOverview} displays the information about the count of each classification of images before and after augmentation.
Original CT images are of size $640 \times 640$ and each image has been resized to $224 \times 224$ to improve the model performance. The data is decomposed into training and testing set as: training set contains (70\%) and testing contains (30\%) of data. Validation is performed on 20\% of the testing data.
\begin{table}[!h]
    \centering
       \caption{Data overview: Count of each class of kidney images before and after augmentation.}
    \begin{tabular}{|l|cc|}
\hline
\multirow{2}{*}{\textbf{Kidney Type}} & \multicolumn{2}{c|}{\textbf{Number of Kindly Images}} \\ \cline{2-3} & \multicolumn{1}{c|}{\textbf{\begin{tabular}[c]{@{}c@{}}Before \\ Augmentation\end{tabular}}} & \textbf{\begin{tabular}[c]{@{}c@{}}After \\ Augmentation\end{tabular}} \\ \hline
Normal & \multicolumn{1}{c|}{5,077}      & 5,141  \\ 
Stone    & \multicolumn{1}{c|}{1,377}  & 3,918  \\ 
Cyst   & \multicolumn{1}{c|}{3,709} & 11,127  \\ 
Tumor & \multicolumn{1}{c|}{2,283} & 6,849  \\ \hline
Total & \multicolumn{1}{c|}{12,446}   & 27,035     \\
\hline
\end{tabular}
    \label{tab:DataOverview}
\end{table}

\subsection{Data augmentation}

\begin{figure}[!ht]
    \centering
\includegraphics[width=0.75\linewidth]{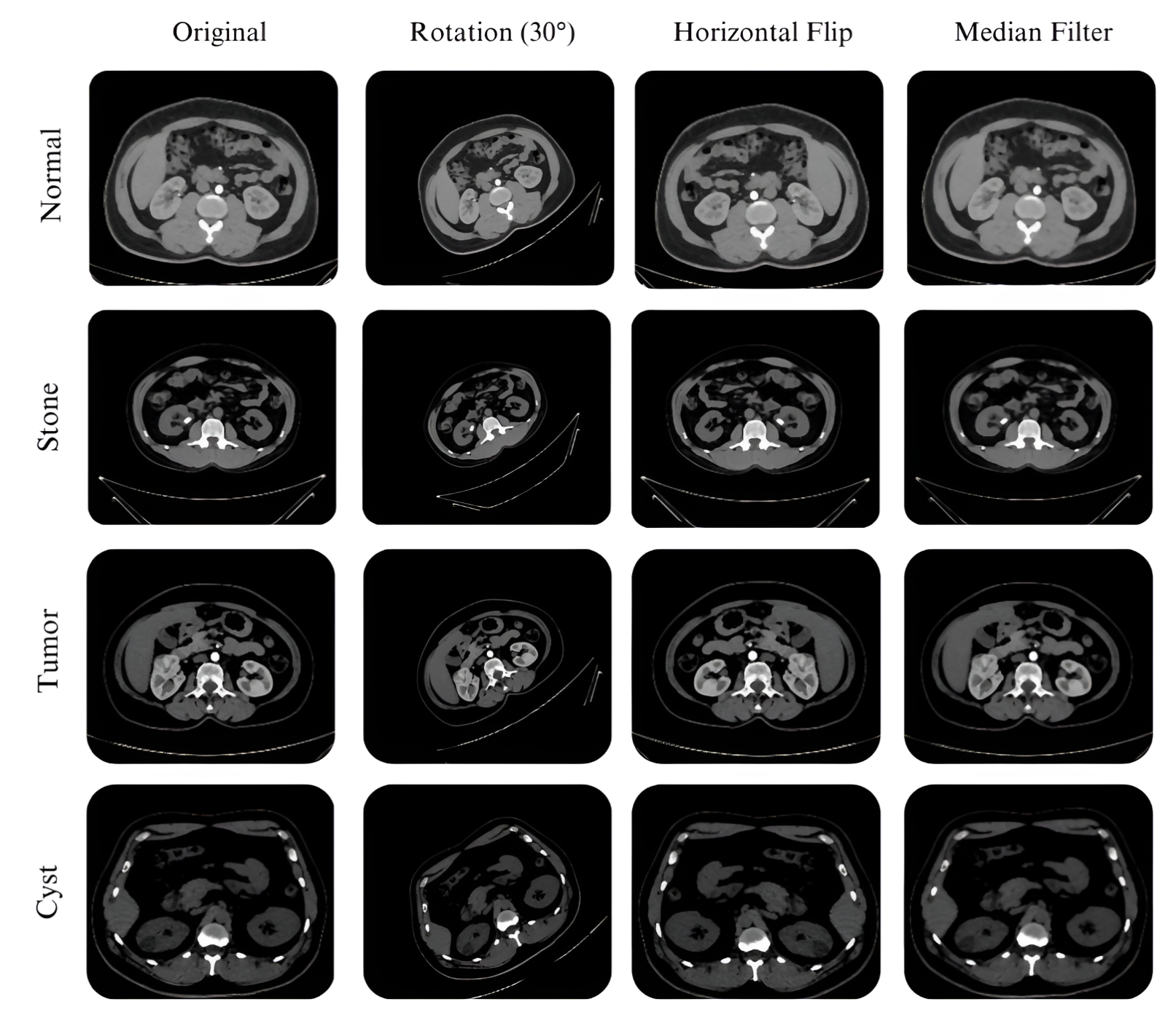} 
\caption{An overview of the augmented data after rotation, flipping, and median filter.}
\label{fig:dataView}
\end{figure}

Data enhancement improves the quality, robustness, and generalizability of machine learning models, especially when data are limited or unbalanced. It involves increasing the size and diversity of the data set by applying transformations such as rotations, translations, scaling, flipping, and cropping to existing data. In this study, flipping images, rotation by 30$^\circ$, and median filtering are used for data augmentation. Figure\ref{fig:dataView} shows the sample of data after augmentation, and the count of each category is given in Table~\ref{tab:DataOverview}.

\section{Architecture of proposed model: Hybrid CNN}

\begin{figure}[!ht]
    \centering
\includegraphics[width=\linewidth]{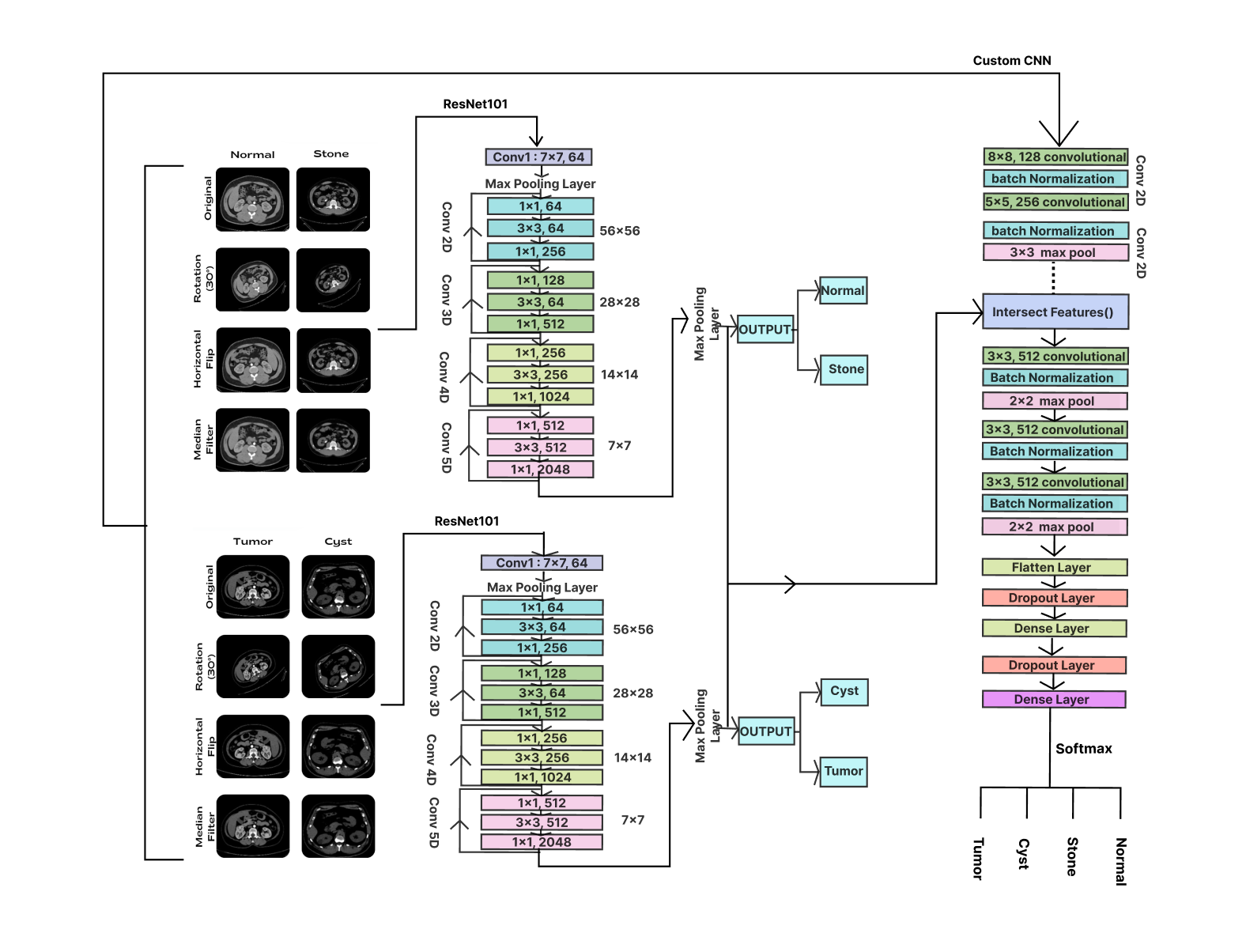} 
\caption{Architecture of the proposed hybrid CNN model.}
\label{fig:architecture}
\end{figure}

The proposed architecture of hybrid CNN shown in Figure~\ref{fig:architecture} illustrates a deep learning architecture to classify kidney images into categories such as normal, stone, tumor and cyst. The architecture has two parts: one it passes input directly to the hybrid CNN model and the other to ResNet101. In Resnet101, the architecture is divided into two specialized branches to address different classification tasks. The first branch handles the classification between normal kidneys and stones, while the second focuses on distinguishing between cysts and tumors. Each branch consists of a series of convolutional layers that extract hierarchical features using various kernel sizes (e.g., 7x7, 3x3, 1x1) and filters, which gradually increase in number (e.g., 64, 256, 512) to capture complex spatial patterns. Batch normalization and max pooling are applied to stabilize training and down-sampling the feature maps, progressively reducing spatial dimensions from 56x56 to 7x7.

The outputs from each branch are passed through classification layers to predict normal vs. stone in one branch and cyst vs. tumor in the other. These features are then merged with the features extracted from hybrid CNN into a unified layer for the final classification. feature mapping filter out the common features with higher similarity. Fully connected layers flatten the feature maps, and a softmax activation function predicts probabilities for all four classes (normal, stone, cyst, tumor). This dual-branch design enables specialized feature extraction for specific abnormalities while providing a comprehensive output. The inclusion of pre-processing, task-specific branches, and hierarchical feature learning makes this architecture efficient for the detection of automated kidney abnormalities in CT images. The proposed model is explained stepwise in the algorithm~\ref{algo}.

\begin{algorithm}
\caption{\textbf{Hybrid CNN}: Feature Extraction and Classification of Kidney CT Images}
\begin{algorithmic}[1]

    \State \textbf{Data Preprocessing}
    \begin{itemize}
        \item  Dataset: labeled kidney images
        \item Normalize pixel values and resize CT images
        \item Apply augmentations (rotation, flipping, and median filter)
    \end{itemize}
  \State \textbf{Feature Extraction}
  \begin{enumerate}
      \item \textbf{ResNet101}
    \begin{itemize}
        \item  Divide the labeled data in two sets: \textit{Normal-Stone} and \textit{Tumor-Cyst}
        \item Use ResNet101 on individual classes to extract features
    \end{itemize}

    \item  \textbf{Custom CNN} 
    \begin{itemize}
        \item Input kidney images to custom CNN
    \item Extract features from images
    \end{itemize}
  \end{enumerate}

    \State \textbf{Training: Hybrid CNN}
    \begin{itemize}
        \item  Pass \textit{Normal-Stone } and \textit{Tumor-Cyst} features extracted from ResNet101 to custom CNN
        \item Perform feature mapping on features extracted from both ResNet101 and custom CNN
        \item Keep the features extracted from both ResNet101 and custom CNN with higher similarity
        \item  Train Hybrid CNN by using cross-entropy loss, batch normalization, max pooling, and Adam optimizer
    \end{itemize}

    \State \textbf{Testing} 
    \begin{itemize}
        \item Pass any kidney CT image to Hybrid CNN model and extract features
        \item Classify the Kidney CT image
    \end{itemize}
    
    \State \textbf{Output}
      \begin{itemize}
        \item Predicted class of kidney: Normal, Stone, Cyst, and Tumor
        \item Evaluate model accuracy: accuracy, precision, recall, and F1 score
    \end{itemize}
\end{algorithmic}
\label{algo}
\end{algorithm}
\subsection{Architecture description}
Table~\ref{tab:output} provides a detailed description of the proposed hybrid CNN architecture, illustrating its layer types, output shapes, and parameter counts. The model begins with a series of Convolutional Layers (Conv2D) to extract spatial features from the input images. These layers progressively increase the number of filters (from 128 to 512) to capture finer and more complex details. Batch Normalization follows each convolutional layer to stabilize and speed up the training process. MaxPooling2D layers are interspersed to downsample the feature maps, reducing their spatial dimensions while retaining essential features.

A specialized Intersect Features function combines relevant features extracted from different models, enhancing the model’s ability to discern fine-grained details. The architecture concludes with Fully Connected (Dense) layers, where the feature maps are flattened into a 1D vector and passed through dense layers to perform high-level representation learning. Dropout layers are included to prevent overfitting by randomly deactivating neurons during training. The final dense layer outputs probabilities for the four target classes: normal, stone, cyst, and tumor. With over 15 million trainable parameters, this architecture effectively balances complexity and efficiency to achieve high classification accuracy.
\begin{table}[!ht]
\centering
\caption{The architecture description of the proposed hybrid CNN model.}
\begin{tabular}{lccll}
\hline
{\color[HTML]{1F1F1F} \textbf{Layer (type)}} & {\color[HTML]{1F1F1F} \textbf{Output Shape}} & {\color[HTML]{1F1F1F} \textbf{\#Param }} \\ \hline
{\color[HTML]{0087FF} conv2d (Conv2D)} & {\color[HTML]{00AF00} (None, 73, 73, 128)} & {\color[HTML]{00AF00} 24,704}  \\ \hline
{\color[HTML]{1F1F1F} batch\_normalization (BatchNormalization)} & {\color[HTML]{00AF00} (None, 73, 73, 128)} & {\color[HTML]{00AF00} 512}  \\ \hline
{\color[HTML]{0087FF} conv2d\_1 (Conv2D)} & {\color[HTML]{00AF00} (None, 73, 73, 256)} & {\color[HTML]{00AF00} 819,456}  \\ \hline
{\color[HTML]{1F1F1F} batch\_normalization\_1 (BatchNormalization)} & {\color[HTML]{00AF00} (None, 73, 73, 256)} & {\color[HTML]{00AF00} 1,024}   \\ \hline
{\color[HTML]{0087FF} max\_pooling2d (MaxPooling2D)} & {\color[HTML]{00AF00} (None, 24, 24, 256)} & {\color[HTML]{00AF00} 0}   \\ \hline
{\color[HTML]{0087FF} conv2d\_2 (Conv2D)} & {\color[HTML]{00AF00} (None, 24, 24, 256)} & {\color[HTML]{00AF00} 590,080} \\ \hline
{\color[HTML]{1F1F1F} batch\_normalization\_2 (BatchNormalization)} & {\color[HTML]{00AF00} (None, 24, 24, 256)} & {\color[HTML]{00AF00} 1,024}  \\ \hline
{\color[HTML]{0087FF} conv2d\_3 (Conv2D)} & {\color[HTML]{00AF00} (None, 24, 24, 256)} & {\color[HTML]{00AF00} 65,792} \\  \hline
{\color[HTML]{1F1F1F} batch\_normalization\_3 (BatchNormalization)} & {\color[HTML]{00AF00} (None, 24, 24, 256)} & {\color[HTML]{00AF00} 1,024}   \\  \hline
{\color[HTML]{0087FF} conv2d\_4 (Conv2D)} & {\color[HTML]{00AF00} (None, 24, 24, 256)} & {\color[HTML]{00AF00} 65,792} \\  \hline
{\color[HTML]{1F1F1F} batch\_normalization\_4 (BatchNormalization)} & {\color[HTML]{00AF00} (None, 24, 24, 256)} & {\color[HTML]{00AF00} 1,024}   \\ \hline

{\color[HTML]{F0AF20} Intersect Features()}   \\ \hline

{\color[HTML]{0087FF} conv2d\_5 (Conv2D)} & {\color[HTML]{00AF00} (None, 24, 24, 512)} & {\color[HTML]{00AF00} 1,180,160}  \\ \hline
{\color[HTML]{1F1F1F} batch\_normalization\_5 (BatchNormalization)} & {\color[HTML]{00AF00} (None, 24, 24, 512)} & {\color[HTML]{00AF00} 2,048}   \\ \hline
{\color[HTML]{0087FF} max\_pooling2d\_1 (MaxPooling2D)} & {\color[HTML]{00AF00} (None, 12, 12, 512)} & {\color[HTML]{00AF00} 0}   \\ \hline
{\color[HTML]{0087FF} conv2d\_6 (Conv2D)} & {\color[HTML]{00AF00} (None, 12, 12, 512)} & {\color[HTML]{00AF00} 2,359,808} \\ \hline
{\color[HTML]{1F1F1F} batch\_normalization\_6 (BatchNormalization)} & {\color[HTML]{00AF00} (None, 12, 12, 512)} & {\color[HTML]{00AF00} 2,048} \\ \hline
{\color[HTML]{0087FF} conv2d\_7 (Conv2D)} & {\color[HTML]{00AF00} (None, 12, 12, 512)} & {\color[HTML]{00AF00} 2,359,808}  \\ \hline
{\color[HTML]{1F1F1F} batch\_normalization\_7 (BatchNormalization)} & {\color[HTML]{00AF00} (None, 12, 12, 512)} & {\color[HTML]{00AF00} 2,048} \\  \hline
{\color[HTML]{0087FF} max\_pooling2d\_2 (MaxPooling2D)} & {\color[HTML]{00AF00} (None, 6, 6, 512)} & {\color[HTML]{00AF00} 0}  \\ \hline
{\color[HTML]{0087FF} conv2d\_8 (Conv2D)} & {\color[HTML]{00AF00} (None, 6, 6, 512)} & {\color[HTML]{00AF00} 2,359,808}  \\ \hline
{\color[HTML]{1F1F1F} batch\_normalization\_8 (BatchNormalization)} & {\color[HTML]{00AF00} (None, 6, 6, 512)} & {\color[HTML]{00AF00} 2,048}  \\ \hline
{\color[HTML]{0087FF} max\_pooling2d\_3 (MaxPooling2D)} & {\color[HTML]{00AF00} (None, 3, 3, 512)} & {\color[HTML]{00AF00} 0}  \\  \hline
{\color[HTML]{1F1F1F} flatten (Flatten)} & {\color[HTML]{00AF00} (None, 4608)} & {\color[HTML]{00AF00} 0}  \\ \hline
{\color[HTML]{0087FF} dense (Dense)} & {\color[HTML]{00AF00} (None, 1024)} & {\color[HTML]{00AF00} 4,719,616}  \\ \hline
{\color[HTML]{1F1F1F} dropout (Dropout)} & {\color[HTML]{00AF00} (None, 1024)} & {\color[HTML]{00AF00} 0} \\ \hline
{\color[HTML]{0087FF} dense\_1 (Dense)} & {\color[HTML]{00AF00} (None, 1024)} & {\color[HTML]{00AF00} 1,049,600}   \\ \hline
{\color[HTML]{1F1F1F} dropout\_1 (Dropout)} & {\color[HTML]{00AF00} (None, 1024)} & {\color[HTML]{00AF00} 0}  \\ \hline
{\color[HTML]{0087FF} dense\_2 (Dense)} & {\color[HTML]{00AF00} (None, 4)} & {\color[HTML]{00AF00} 4,100}  \\ \hline
\end{tabular}
\label{tab:output}
\end{table}

\subsection{Testing of the model}
\begin{figure}[!ht]
    \centering
\includegraphics[width=0.85\linewidth]{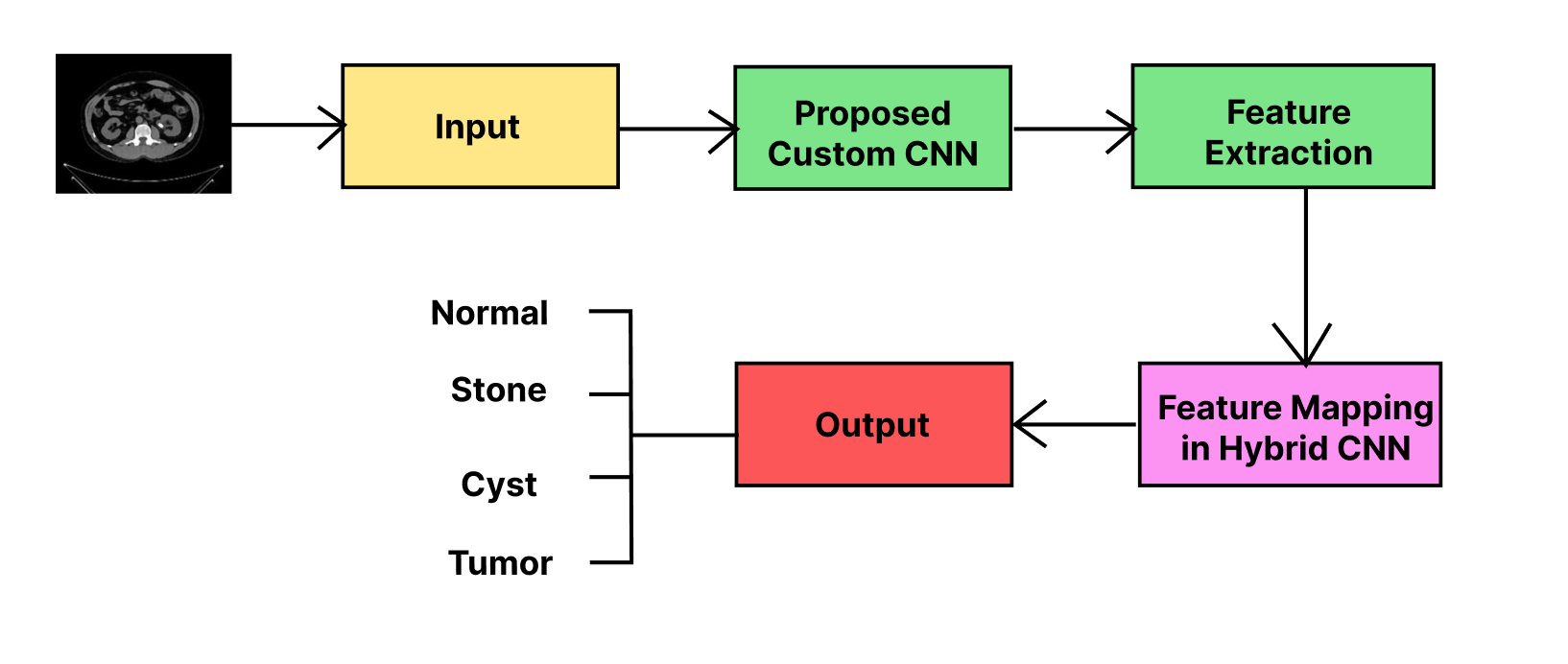} 
\caption{Schematic outline of hybrid CNN testing.}
\label{fig:testing}
\end{figure}

Figure~\ref{fig:testing} illustrates the workflow of a hybrid CNN-based model designed for the classification of kidney CT images. Initially, the CT image of the input kidney is processed by the proposed custom CNN to extract relevant features. These features are then refined and passed through a feature extraction process to generate meaningful data representations. The extracted features are subsequently combined using a feature mapping module within the hybrid CNN architecture. This fusion process ensures the integration of diverse set of features for robust classification. Finally, fused features are classified into one of four categories: normal, stone, cyst, or tumor. 

\section{Experimental setup}
This section elaborates the experimental setup that includes the configuration and implementations of the proposed hybrid CNN along with the evaluation metrics.

\subsection{Configuration}
 Table~\ref{tab:hyperparameters} shows all the pre-initialized parameters used for hybrid CNN. It specifies that the model is trained using 5 epochs and a batch size of 32. The learning rate is set to 0.001, which dictates the speed of weight updates during training. The stride, which controlled the size of the step during convolution, is set to 1, ensuring fine-grained feature extraction. The first layer of the model comprises 128 kernels, each applying a convolution operation to extract features, and uses the ReLU activation function for non-linearity. The dropout rate of 0.5 is applied to prevent overfitting by randomly deactivating half the neurons during training. The model has 15,605,124 trainable parameters and 6,400 non-trainable parameters, emphasizing its complexity. On average, it required 132 seconds per epoch for training. These details highlight the design choices aimed at optimizing the hybrid CNN performance.

\begin{table}[!ht]
\centering
\caption{Hyperparameters used in the proposed model.}
\begin{tabular}{ll}
\hline
\textbf{Hybrid CNN Hyperparameters} & \textbf{Details} \\ \hline
Number of epochs                                     & 5                \\
Batch size                                      & 32               \\
Learning rate                                   & 0.001            \\
Stride                                          & 1                \\
Number of kernels in different layers           & 128, 256 and 512            \\
Activation                                      & ReLu             \\
Dropout rate                                    & 0.5              \\
Number of trainable parameters                  & 15,605,124         \\
Number of non trainable parameters              & 6,400             \\
 \hline
\end{tabular}
\label{tab:hyperparameters}
\end{table}
\subsection{Computational setup}
The  model operates on a system equipped with a T4 GPU, 12.7 GB of system RAM, 15.0 GB of GPU RAM, and 112.6 GB of disk space, providing a robust computational environment. The model runs efficiently for up to 10 epochs, using these resources optimally. Beyond this threshold, it fully utilizes the available system and GPU resources, highlighting the system's computational capacity and the model's demand for memory and processing power during extended training. 
\subsection{Evaluation metrics}

The following properties are evaluated to measure the performance of the model.

\begin{enumerate}
    \item \textbf{Precision:} Precision measures the accuracy of positive predictions and is caluclated as
    \begin{equation}
        \text{\textit{Precision}} = \frac{TP}{TP + FP}
    \end{equation}
    where $TP$ is true positive and $FP$ is false positive.
 \item \textbf{Recall:} Recall, or sensitivity, measures the ability of a model to find all true positives cases and is calculated as 
     \begin{equation}
        \text{\textit{Recall}} = \frac{TP}{TP + FN}
    \end{equation}
   
where $FN$ is false negative.
 \item \textbf{F1-Score:} The F1 score is the harmonic mean of precision and recall and is calculated as

     \begin{equation}
       \text{\textit{F1-Score}} = 2 \times \frac{\text{Precision} \times \text{Recall}}{\text{Precision} + \text{Recall}}
    \end{equation}
 \item \textbf{Accuracy:} Accuracy measures the overall correctness of the model and is calculated as

     \begin{equation}
       \text{\textit{Accuracy}} = \frac{TP + TN}{TP + TN + FP + FN}
    \end{equation}
where $TN$ is true negative.
\end{enumerate}

Table~\ref{tab: classification}  provides a classification report for the performance of a model over five epochs, detailing various evaluation metrics. The precision increased from 74.83\% in Epoch 1 to 99.73\% in Epoch 5, with precision, recall, and F1 scores reaching perfect values (100\%) from Epoch 3 onward. The average standard deviation decreased significantly, from 0.0942 in Epoch 1 to 0.0012 in Epoch 5, reflecting increased stability in predictions. 

\begin{table}[!ht]
\centering
\caption{Training hybrid CNN: Performance metrics of the model for classification of kidney as normal, stone, cyst, and tumor for each epoch.}
\begin{tabular}{cccccccc}
\hline
\textbf{Epoch} & \textbf{Accuracy (\%)} & \textbf{Precision (\%)} & \textbf{Recall (\%)}  & \textbf{F1-Score (\%)} & \textbf{Std Dev} \\ \hline
1              & 74.83             & 96.71              & 96.59                             & 96.48             & 0.0942            \\
2              & 94.78             & 99.84              & 99.84                            & 99.84             & 0.0219            \\
3              & 98.58             & 100                & 100                              & 100               & 0.0061            \\
4              & 99.45             & 100                & 100                                 & 100               & 0.0023            \\
5              & 99.73             & 100                & 100                                & 100               & 0.0012       \\ \hline    
\end{tabular}
\label{tab: classification}
\end{table}
\section{Results and discussion}

\subsection{Accuracy and loss}
\begin{figure}[!ht]
    \centering
\includegraphics[width=0.48\linewidth]{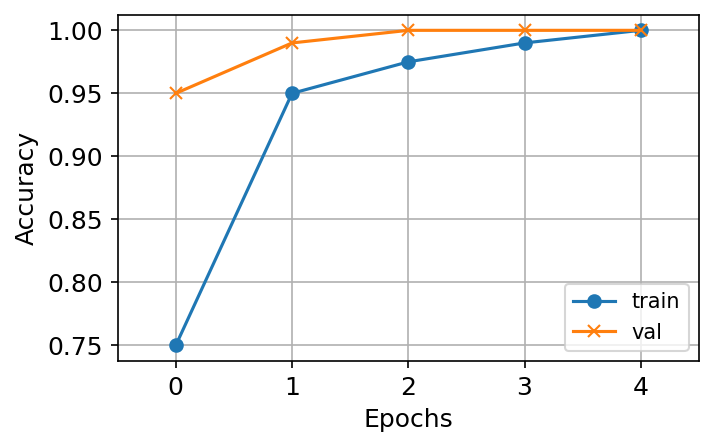} 
\includegraphics[width=0.48\linewidth]{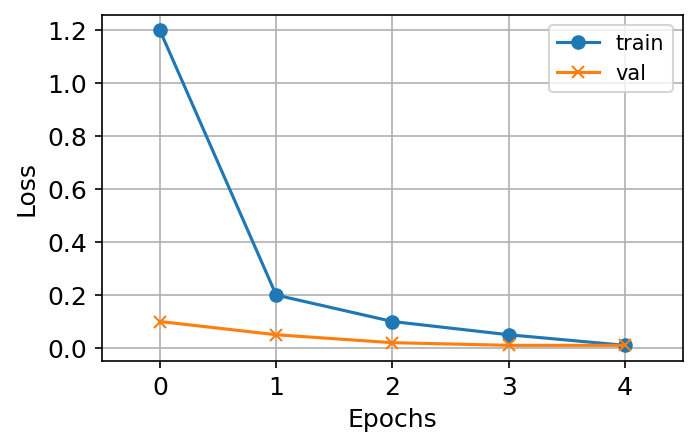} 
\caption{(\textit{left}) Model accuracy trend. (\textit{right}) Model loss trend.}
\label{fig:Accuracy_Loss}
\end{figure}

Figure~\ref{fig:Accuracy_Loss} (left) illustrates the accuracy trends for the training and validation datasets in five epochs during model training. Training accuracy starts around 74\% in Epoch 1 and rapidly increases, reaching nearly 100\% by Epoch 3, maintaining this high performance through subsequent epochs. This demonstrates that the model quickly learned the training data with high precision. However, validation accuracy begins higher than training accuracy at approximately 95\% in Epoch 1 and achieves 100\% by Epoch 3, remaining stable through Epochs 4 and 5. The steady and close alignment with the training accuracy indicates that the model generalizes well and avoids overfitting. The overlapped accuracy values after Epoch 3 signify a well-trained model with balanced performance on both datasets.

Similarly, Figure~\ref{fig:Accuracy_Loss} (right) shows the loss trends for both the training and validation data sets in five epochs during model training. Training loss starts high at around 0.65 in Epoch 1 and decreases sharply to near 0 by Epoch 3. This indicates that the model effectively minimizes error in the training dataset as the training progresses. However, validation loss starts at around 0.2 in Epoch 1 and also decreases steadily to near 0 by Epoch 3, maintaining this minimal loss in subsequent epochs. The low and stable validation loss suggests that the model generalizes well to unseen data and avoids overfitting. Rapid convergence of both training and validation losses to near-zero values highlights efficient learning of the model and excellent performance on both datasets at the end of training.

\subsection{Comparative analysis}

To test the performance of the model, a test sample of 3,734 CT images is used. Table~\ref{tab:comparison} shows the comparison of hybrid CNN with pre-trained ResNet101 and base CNN models. Among the three, hybrid CNN outperforms the others, achieving 100\% accuracy, precision, recall, and F1 score, demonstrating its ability to classify all images perfectly without false positives or negatives. ResNet101 also performs strongly, with an accuracy of 99.81\%, precision of 99.75\%, recall of 99.50\%, and an F1 score of 99.75\%, while the base CNN is significantly short, with an accuracy of 77.39\%, precision of 78.10\%, recall of 71.72\%, and an F1 score of 72.80\%. The training time is longest for the hybrid CNN due to its complexity, compared to ResNet101 and base CNN. However, the hybrid CNN compensates for this with the fastest testing time compared to other two. This indicates that the hybrid CNN effectively combines the strengths of pre-trained ResNet101 and the custom CNN, resulting in exceptional classification performance with practical efficiency.
\begin{table}[!ht]
\centering
\caption{Testing the performance of hybrid CNN with base CNN and pre-trained ResNet101.}
\begin{tabular}{lcccccccc}
\hline
\textbf{Models} & \textbf{\begin{tabular}[c]{@{}c@{}}Accuracy \\ (\%)\end{tabular}} & \textbf{\begin{tabular}[c]{@{}c@{}}Precision \\ (\%)\end{tabular}} & \textbf{\begin{tabular}[c]{@{}c@{}}Recall \\ (\%)\end{tabular}} &  \textbf{\begin{tabular}[c]{@{}c@{}}F1-Score \\ (\%)\end{tabular}} & \textbf{\begin{tabular}[c]{@{}c@{}}Training \\ Time (hrs)\end{tabular}} & \textbf{\begin{tabular}[c]{@{}c@{}}Testing \\ Time (sec)\end{tabular}} \\ \hline


ResNet101  & 99.81   & 99.75   & 99.50  & 99.75   & 0:47:55   & 31  \\


Hybrid CNN     & 100   & 100   & 100   & 100   & 1:25:02   & 23   \\ \hline                                                          
\end{tabular}

\label{tab:comparison}
\end{table}
\subsubsection{Confusion matrix}
We assessed the performance of the hybrid CNN by analyzing its confusion matrix in comparison with ResNet101. The confusion matrix provides a two-dimensional representation where rows represent the actual class labels (Cyst, Normal, Stone, Tumor) and columns represent the predicted labels by the model. The ResNet101 model demonstrated high precision across all four classes, accurately classifying 1,017 out of 1,018 cysts, 1,531 out of 1,534 normals, 423 out of 428 stones, and 648 out of 654 tumors. The highest misclassification occurred with four stones being incorrectly predicted as cysts, but overall, the true positive rates were exceptional, showcasing robust feature extraction and classification capabilities. When comparing the hybrid CNN’s performance to ResNet101, the hybrid CNN achieved perfect classification accuracy. It correctly identified 1,118 cases of Cyst, 1,534 cases of Normal, 428 cases of Stone, and 654 cases of Tumor, with zero misclassifications.
\begin{figure}[!ht]
    \centering 
\includegraphics[width=\linewidth]{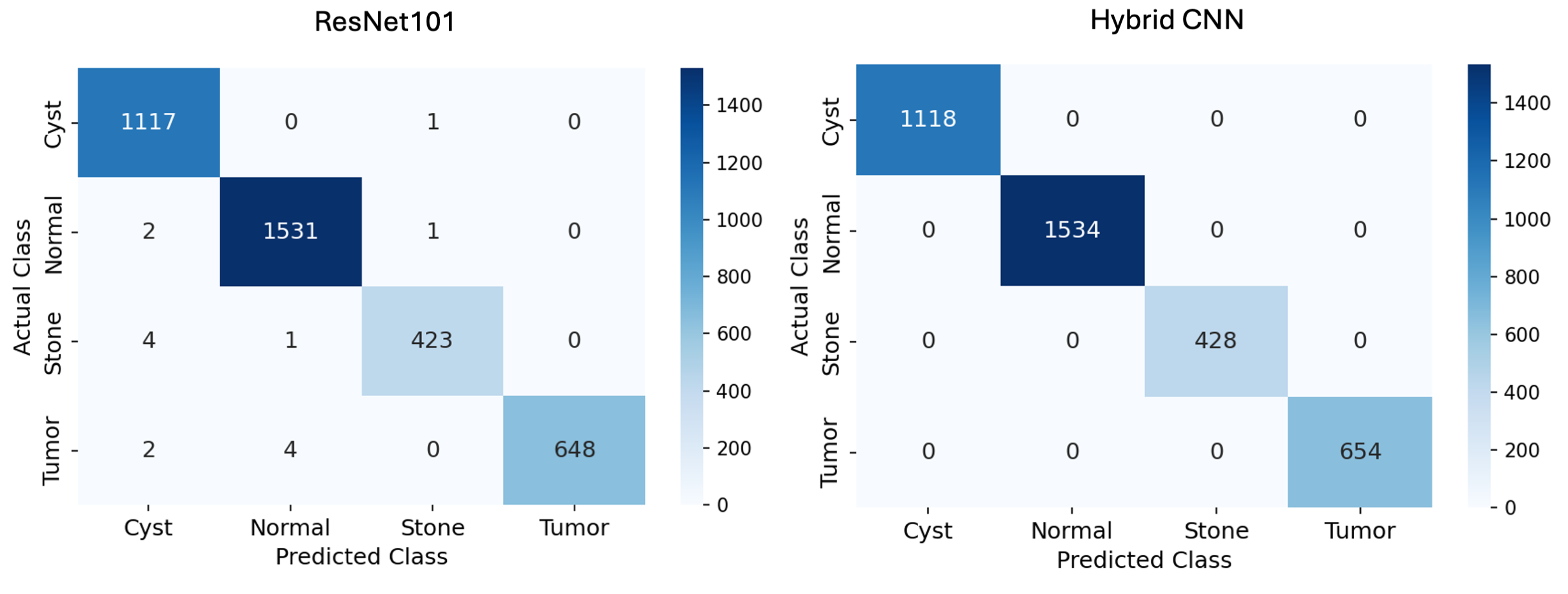} 
\caption{Comparative analysis of confusion matrix. (\textit{left}) ResNet101 and (\textit{right}) Hybrid CNN.}
\label{fig:ConfusionMatrix}
\end{figure}


 \subsubsection{Principal component analysis (PCA)}
\begin{figure}[!ht]
    \centering 
\includegraphics[width=\linewidth]{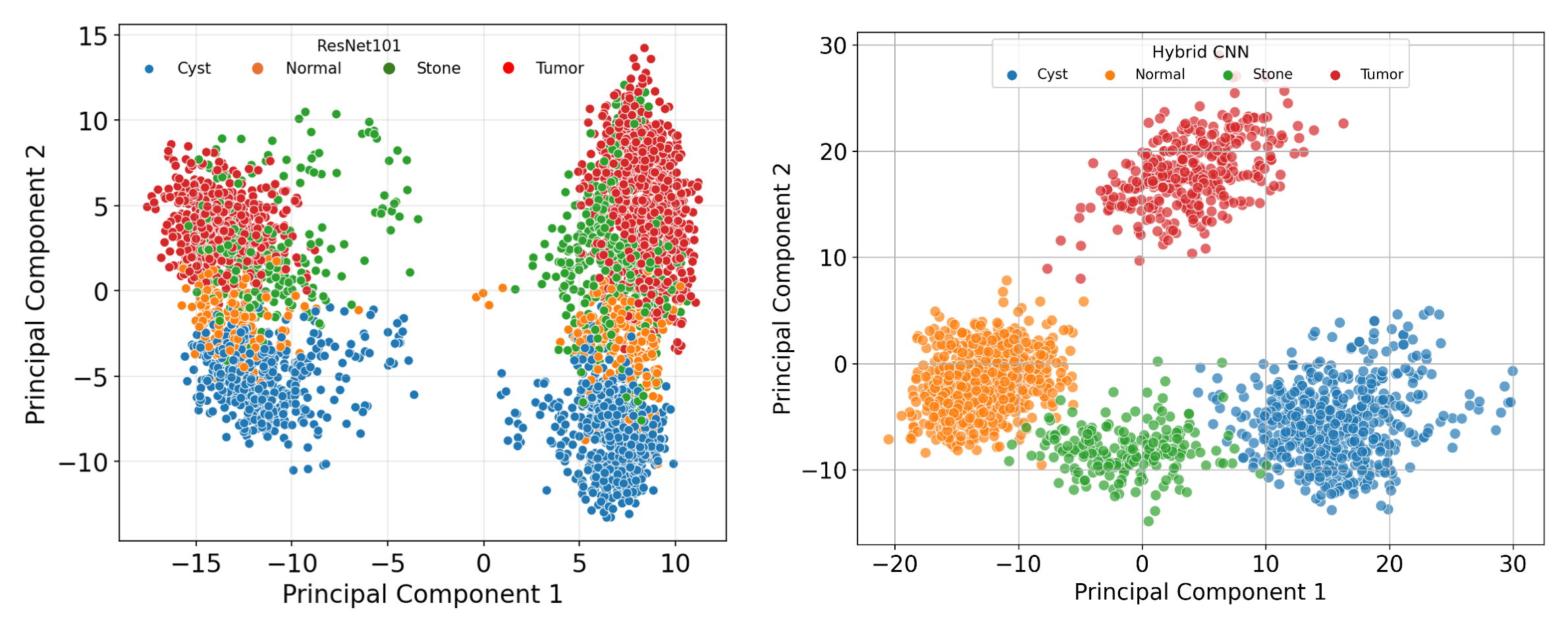} 
\caption{Comparative analysis of confusion matrix. (\textit{left}) ResNet101 and (\textit{right}) Hybrid CNN.}
\label{fig:PCA}
\end{figure}

 Figure~\ref{fig:PCA} shows the confusion matrix of the ResNet101 model. 

Principal Component Analysis (PCA) is used to reduce the high-dimensional feature space from CT images to two dimensions, while preserving the maximum variance. PCA computes the eigenvectors of the covariance matrix, which represent directions of maximum variance in the data. This allows visualization of the distribution of four classes: healthy kidney, kidney stone, cyst, and tumor. The first two principal components, PC1 and PC2 for  ResNet101, and proposed hybrid CNN, are shown in Figure~\ref{fig:PCA}. 

In both PCA's, each point represents a CT image, colored according to its class: Normal (orange), Stone (green), Cyst (blue), and Tumor (red).
The plot reveals a clear separation between classes. Normal kidney samples primarily cluster in the lower region, cyst samples form distinct clusters in the upper regions, and Stone and Tumor samples overlap in the intermediate regions. Kidney stone samples and tumor samples overlap, probably due to shared features such as irregular shapes or intensity patterns.
    

PCA for ResNet1010 shows a mirrored pattern along the PC1 axis. The left half $(PC1 < 0)$ likely represents one subgroup, while the right half $(PC1 > 0)$ represents another. Almost mirror images in scatter plots might occur because PCA relies on eigenvectors and eigenvalues that define directions in space, but do not have inherent orientations. Eigenvectors are normalized to have unit length, but their signs are arbitrary. For example, if $\left[ v,v,v\right]$ is an eigenvector, $\left[-v,-v,-v\right]$ is also valid because both represent the same direction. This mirroring could also be due to the following reasons:

\begin{enumerate}
    \item \textit{Non-linear or Multi-modal Data Structure}: Variations in CT images or differences within each class (e.g., cyst size, stone composition, tumor location) could cause this split.
\item \textit{Data Augmentation:} Techniques like rotation, flipping, mirroring up/down, etc. may create separate groups, as the augmented features are treated as distinct by PCA.
\end{enumerate}
The mirroring does not affect the interpretation of PCA because it does not change the variance explained or the relative distances between points in the reduced space. The PCA plots 1 and 2 demonstrates ResNet101 ability to separate healthy kidneys from pathological conditions. However, the overlap between Tumor and Stone classes highlights challenges in differentiating these conditions based on the current feature set. This suggests that additional features or a more refined model could improve the classification accuracy for overlapping classes. 
The proposed hybrid CNN model (Figure~\ref{fig:PCA}(right)) shows a clearer separation of the four classes without the mirroring effect, indicating that the hybrid CNN model resolves the subgroup issue by filtering common features.

\section{Conclusion}
To conclude, the hybrid CNN model excelled in classifying kidney CT images, achieving 100\% accuracy, precision, recall, and F1 score. By combining ResNet101 features and a custom CNN via feature fusion, the model captured diverse feature representations, significantly outperforming standalone models. 
This dual-branch architecture facilitated robust feature extraction and mapping, enabling precise differentiation across all kidney conditions. The study underscores the potential of the model to optimize clinical workflows, improve diagnostic accuracy, and support early intervention strategies, thus significantly improving patient care and outcomes.

\subsection{Limitations}

Despite its high performance, the proposed model has some limitations. First, the training time is longer compared to individual models due to the complexity of the hybrid architecture. Second, the model relies on a relatively large, well-annotated dataset, which may not always be available in real-world scenarios.  
\bibliographystyle{cas-model2-names}
\bibliography{cas-refs}

\section*{Ethical Statement}
The authors declare that the study did not involve human participants or animals.

\section*{Data Availability Statement}
The data used in this study is available on \href{https://www.kaggle.com/datasets/srinivasbece/kindey-stone-dataset-splitted}{https://www.kaggle.com/datasets/srinivasbece/kindey-stone-dataset-splitted}.  
The data is also available through the corresponding author on reasonable request.

\section*{Author Contributions}
KS conceptualized the research problem, conducted the analysis, wrote and reviewed the manuscript.  
ZU conceptualized the research problem, performed analysis, wrote and reviewed the manuscript.  
AW collected the data and carried out the analysis.  
DG designed the analysis framework and contributed to figure preparation.

\section*{Funding}
No funding is available.

\section*{Competing Interests}
The authors declare that they have no conflict of interest.

\end{document}